\documentclass[11pt,a4paper]{article}
\usepackage{jcappub}

\usepackage[top=2.8cm, bottom=2.8cm, left=2.4cm, right=2.4cm]{geometry}
\usepackage[utf8]{inputenc}  
\usepackage[T1]{fontenc}     
\usepackage{amsmath,amssymb,lmodern} 
\usepackage{graphicx}
\usepackage{hyperref}
\usepackage{color}

\newcommand{\ba}{\begin{eqnarray}}
\newcommand{\ea}{\end{eqnarray}}
\newcommand{\be}{\begin{equation}}
\newcommand{\ee}{\end{equation}}
\newcommand{\bea}{\begin{eqnarray}}
\newcommand{\eea}{\end{eqnarray}}
\newcommand{\beq}{\begin{equation}}
\newcommand{\eeq}{\end{equation}}
\newcommand{\beqar}{\begin{eqnarray}}
\newcommand{\eeqar}{\end{eqnarray}}
\newcommand{\beqars}{\begin{eqnarray*}}
\newcommand{\eeqars}{\end{eqnarray*}}
\newcommand{\bc}{\begin{center}}
\newcommand{\ec}{\end{center}}
\newcommand{\ben}{\begin{enumerate}}
\newcommand{\een}{\end{enumerate}}
\newcommand{\bit}{\begin{itemize}}
\newcommand{\eit}{\end{itemize}}
\newcommand{\bw}{\begin{widetext}}
\newcommand{\ew}{\end{widetext}}
\newcommand{\bcl}{\begin{columns}}
\newcommand{\ecl}{\end{columns}}

\newcommand{\dd}{\mbox{d}}

\begin{document}

\subheader{PI/UAN-2016-595FT}
\title{On the 4D generalized Proca action for an Abelian vector field}

\author[a]{Erwan Allys,}
\emailAdd{allys@iap.fr}
\affiliation[a]{Institut d’Astrophysique de Paris, UMR 7095, \\
UPMC Universit\'e Paris 6 et CNRS,\\
98 bis boulevard Arago, 75014 Paris, France}

\author[b]{Juan P. Beltr\'an Almeida,}
\emailAdd{juanpbeltran@uan.edu.co}
\affiliation[b]{Departamento de  F\'isica,  Universidad  Antonio Nari\~no, \\
Cra 3 Este \# 47A-15, Bogot\'a D.C. 110231, Colombia}

\author[a,c]{Patrick Peter}
\emailAdd{peter@iap.fr}
\affiliation[c]{Institut Lagrange de Paris,\\
UPMC Universit\'e Paris 6 et CNRS,\\
Sorbonne Universit\'es, Paris, France}

\author[d,e,f]{and Yeinzon Rodr\'iguez}
\emailAdd{yeinzon.rodriguez@uan.edu.co}
\affiliation[d]{Centro de Investigaciones en Ciencias B\'asicas y Aplicadas, Universidad Antonio Nari\~no, \\
Cra 3 Este \# 47A-15, Bogot\'a D.C. 110231, Colombia}
\affiliation[e]{Escuela  de  F\'isica,  Universidad  Industrial  de  Santander, \\
Ciudad  Universitaria,  Bucaramanga  680002,  Colombia}
\affiliation[f]{Simons Associate at The Abdus Salam International Centre for Theoretical Physics, \\
Strada Costiera 11, I-34151, Trieste, Italy}

\date{\today}

\abstract{
We summarize previous results on the most general Proca theory in 4 dimensions containing only first-order derivatives in the vector field (second-order at most in the associated Stückelberg scalar) and having only three propagating degrees of freedom with dynamics controlled by second-order equations of motion.  Discussing the Hessian condition used in previous works, we conjecture that, as in the scalar galileon case, the most complete action contains only a finite number of terms with second-order derivatives of the Stückelberg field describing the longitudinal mode, which is in agreement with the results of JCAP {\bf 1405}, 015 (2014) and Phys. Lett. B {\bf 757}, 405 (2016) and complements those of JCAP {\bf 1602}, 004 (2016).  We also correct and complete the parity violating sector, obtaining an extra term on top of the arbitrary function of the field $A_\mu$, the Faraday tensor $F_{\mu \nu}$ and its Hodge dual $\tilde{F}_{\mu \nu}$.}

\maketitle
\section{Introduction}

Along the line of modifying gravity in a scalar-tensor way, many
proposals have been made to write down theories whose dynamics stem from
second order equations of motion for both the tensor and the scalar
degrees of freedom
\cite{Nicolis:2008in,Deffayet:2009wt,Deffayet:2009mn,Deffayet:2011gz,deRham:2011by},
thus generalizing an old proposal \cite{Horndeski:1974wa}; such theories have been dubbed Galileons. 
The obvious next move consists in obtaining a similar general action for a vector field \cite{Horndeski:1976gi} 
(see also in Refs. \cite{Zumalacarregui:2013pma,Gleyzes:2014dya,Langlois:2015cwa,Langlois:2015skt}), 
thereby forming the vector Galileon case \cite{Heisenberg:2014rta,Tasinato:2014eka}, which 
was investigated thoroughly \cite{EspositoFarese:2009aj,Hull:2014bga,Khosravi:2014mua,Charmchi:2015ggf,Hull:2015uwa,Li:2015vwa,Li:2015fxa,Heisenberg:2016eld}. 
Demanding U(1) invariance led to a no-go theorem \cite{Deffayet:2013tca} which 
can be by-passed essentially by dropping the U(1) invariance hypothesis. Cosmological
implications of such a model can be found e.g. in Refs.~\cite{BeltranJimenez:2010uh,Barrow:2012ay,Jimenez:2013qsa,Tasinato:2013oja,Tasinato:2014mia,Jimenez:2015caa,Gao:2016vtt,Heisenberg:2016dkj,DeFelice:2016yws,Nieto:2016gnp,Chagoya:2016aar,DeFelice:2016uil,Heisenberg:2016wtr}.

Recent papers \cite{Heisenberg:2014rta,Allys:2015sht,Jimenez:2016isa}
have derived the most general action containing a vector field, with
different conclusions as to the number of possible terms given the
underlying hypothesis. In Refs.~\cite{Heisenberg:2014rta,Jimenez:2016isa},
the Lagrangian was built from contractions of derivative terms with
Levi-Civita tensors, whereas Ref.~\cite{Allys:2015sht} used a more systematic
approach based on the Hessian condition. It appears that a consensus has finally been
reached, suggesting only a finite number of terms in the theory, all of
them being given in an explicit form. To describe this consensus and
complete the discussion, we examine in the present paper an alternative
explanation for the presence of, allegedly, only a finite number of
terms in the generalized Proca theory, using the tools developed in
Ref.~\cite{Allys:2015sht} where the infinite series of terms was
conjectured. This discussion also allows us to
compare the systematic procedure used in Ref.~\cite{Allys:2015sht} with the
construction based on Levi-Civita tensors of
Refs.~\cite{Heisenberg:2014rta,Jimenez:2016isa}.  We then summarize these
previously obtained results and settle the whole point in as definite a
manner as possible.

Focusing on the parity violating sector of the model, not thoroughly
investigated in Refs. \cite{Heisenberg:2014rta,Jimenez:2016isa}, 
certain terms obtained in Ref.~\cite{Allys:2015sht}
should not appear according to the abovementioned discussion.
Indeed, we show that, because of an identity not taken into
account in Ref.~\cite{Allys:2015sht}, those unexpected parity-violating terms are
either merely vanishing or can be combined into a simple scalar formed
with the field $A^\mu$, the Faraday tensor $F^{\mu\nu}$ and its Hodge
dual $\tilde{F}^{\mu\nu}$. 
This closes the gap, hence providing an
even firmer footing to the conjecture according to which the most general
theory is in fact given by Eq.~\eqref{TotalLag}, which is, up to a new term uncovered in this paper, 
Eq.~(12) of Ref.~\cite{Jimenez:2016isa} in
Minkowski space, or Eq.~(28) in an arbitrary curved spacetime.

In Sec.~\ref{PartStateOfTheArt}, we summarize the results previously
obtained, together with the associated investigation procedures. We then
present the generic structure in Sec.~\ref{conj}, emphasizing how it
permits an automatic implementation of the Hessian condition, and argue
that the number of acceptable Lagrangian structures satisfying the usual
physical requirements is finite, up to arbitrary functions. Splitting
the possible terms into parity conserving and violating contributions,
we motivate our conclusion in Secs.~\ref{ParCons} and \ref{ParViol},
resolving the apparent disagreement between the present conjecture and
the conclusions of a previous work \cite{Allys:2015sht}; we conclude in
Sec.~\ref{PartFinalModel} by explicitly writing down the final 4D vector
action.

\section{Present status}
\label{PartStateOfTheArt}

Let us first introduce the vector theory, the hypothesis and results
obtained thus far. We assume in what follows the Minkowski metric to
take the form $g^{\mu \nu}=\eta^{\mu \nu}=\rm{diag}(-1,+1,+1,+1)$ and
set $(\partial \cdot A) \equiv \partial_\mu A^\mu$ and $X=A_\mu A^\mu$
for simplification and notational convenience.

\subsection{Generalized Abelian Proca theory}

One seeks to generalize Proca theory, namely that stemming from the action
\begin{equation}
\mathcal{S}_\mathrm{Proca}  = \int \mathcal{L}_\mathrm{Proca}\,\dd^4x 
= \int \left( -\frac14 F_{\mu\nu} F^{\mu\nu} + \frac12 m_A^2 X \right)\,\dd^4x \,,
\end{equation}
with $A^\mu$ being a massive vector field, not subject to satisfy a U(1)
invariance, and $F_{\mu\nu} \equiv \partial_\mu
A_\nu-\partial_{\nu}A_{\mu}$ being the associated Faraday tensor. The
generalization of this action can be made by considering all ``safe''
terms containing the vector field and its first derivative. To explicit
what safe means in this context, one
decomposes the field into a scalar $\pi$ and pure vector $\bar{A}$ parts
according to
\begin{equation}
A_\mu=\partial_\mu \pi + \bar{A}_\mu \ \ \ \hbox{or} \ \ \ A = \dd \pi +
\bar{A} \,,
\end{equation}
where $\pi$ is commonly referred to as the Stückelberg field, and
$\bar{A}^\mu$ is divergence-free. One then demands the equations of
motion for $A^\mu$, and for both $\pi$ and $\bar{A}^\mu$, are
second order, and that the Proca field propagates only three degrees of
freedom \cite{deRham:2010ik}. These conditions are discussed in full depth in
Refs.~\cite{Heisenberg:2014rta,Allys:2015sht,Jimenez:2016isa}. The first
condition ensures that the model can be stable \cite{ostro,Boulware:1973my,Woodard:2006nt}, while
the second stems from the fact that a massive field of spin $s$
propagates $2s+1$ degrees of freedom.

Note that the scalar field will appear in two different parts, one
containing only the Stückelberg field itself, and one containing also the
pure vector contribution $\bar{A}$. Examining the decoupling limit of the
theory, one recovers for the pure scalar part of the Lagrangian the exact
requirements of the Galileon theory
\cite{Nicolis:2008in,Deffayet:2009wt,Deffayet:2009mn,Deffayet:2011gz},
and so this part of the Lagrangian must reduce to this well-studied class
of model.

\subsection{Investigation procedures}
\label{PartInvestigationProcedure}

Two different but equivalent procedures have been devised to write down
the most general theory sought for. The first, originally proposed and explained in
Refs.~\cite{Heisenberg:2014rta,Jimenez:2016isa}, consists in a systematic
construction of scalar Lagrangians in terms of contractions of two Levi-Civita
tensors with derivatives of the vector field. This permits an easy
comparison with the Galileon theory, as the same structure automatically
ensues. The condition that only three degrees of freedom propagate is
then verified on the relevant terms.

The second procedure, put forward in Ref.~\cite{Allys:2015sht}, works somehow
the other way around by systematically constructing all possible scalar
Lagrangians propagating only three degrees of freedom. To achieve this
requirement, a condition on the Hessian of the Lagrangian $\mathcal{L}$
(or each independent such Lagrangian) considered, namely
\begin{equation}
\mathcal{H}^{\mu\nu}=\frac{\partial^2 \mathcal{L}}{\partial(\partial_0
A_\mu)\partial(\partial_0A_\nu)} \,,
\label{Hessian}
\end{equation}
is imposed. As discussed in Ref. \cite{Allys:2015sht}, in order that the
timelike component of the vector field does not propagate in non trivial
theories, the components $\mathcal{H}^{0\mu}$ must vanish. All possible
terms satisfying this constraint are considered at each order.

There are two crucial points concerning the latter method that still need to
be checked once the terms satisfying all other requirements have been
obtained: not only they must reduce to the scalar galileon Lagrangians in
the pure scalar sector, but they must imply second class constraints.
Moreover, given that it is a systematic expansion in terms of scalars
built out of vectors with derivatives, one must make sure they are not
either identically vanishing or mere total derivatives. In other words,
although the method ensures that all possible terms will be found, they
are somehow too numerous and there may remain some redundancy that must
be tracked down and eliminated.


\section{Generic structure}
\label{conj}

As all contractions between vector derivatives and $\delta$ and $\epsilon$
can always be written in terms of $\epsilon$ only, a complete basis for
expanding the general category of Lagrangians of interest is made up with
terms of the form
\begin{equation}
\mathcal{T}^i_N = \underbrace{\epsilon^-{}_- \epsilon^-{}_- \cdots}_N \
 \partial_\centerdot A_\centerdot \partial_\centerdot A_\centerdot\cdots \,,
\label{GenForm}
\end{equation}
where indices appearing in the field derivatives are contracted only
with corresponding indices in the Levi-Civita tensors, the remaining 
indices being contracted possibly in between Levi-Civita tensors in
such a way as to yield a scalar. Each index $i$ reflects the fact that
there can be more than one way to contract the $N$ Levi-Civita 
tensor to form a scalar.
These terms form a complete basis for the Lagrangians containing an arbitrary
number of field derivatives.

The general Lagrangian will then be of the form
\begin{equation}
\mathcal{L} = \sum_{i,N} f^i_N(X) \mathcal{T}^i_N \,,
\label{GenLag}
\end{equation}
where we consider only prefactors that are functions of $X$: one could
envisage contracting a vector field itself with the derivative terms
involved in Eq. \eqref{GenLag}, but that would lead to an equivalent
basis up to integrations by parts \cite{Allys:2015sht}\footnote{We have found
one special case, discussed below Eq.~\eqref{FS}, for which the total
derivative of the integration by parts would actually vanish for
symmetry reasons; we included and discussed this special term in our
final form of the action.}. When written in terms of the Stückelberg
field only, i.e. setting $A_\mu \to \partial_\mu\pi$, and restricting
attention to $N=2$, Eq. \eqref{GenLag} automatically yields the subclass
of the generalised galileon theory \cite{Deffayet:2009mn} containing
only derivatives of the scalar field\footnote{The full generalized
galileon theory is recovered if one also makes the replacement $f^i_N(X)
\to f^i_N(\pi,\partial \pi)$.}.

The terms thus built in Eq. \eqref{GenForm} now fall into two distinct
categories, depending on how they behave under a U(1) gauge
transformation. Those invariant under such transformations contracts
all field derivative indices to one and only one Levi-Civita tensor, i.e.
they take the form
$$\epsilon^{\mu\nu -} \epsilon^{\rho\sigma -} \cdots \partial_\mu A_\nu
\partial_\rho A_\sigma\cdots \,,$$ which can all be equivalently expressed as functions of
scalar invariants made out of the Faraday tensor  $F_{\mu\nu}$ and its
Hodge dual $\tilde{F}^{\mu\nu}=\frac{1}{2}\epsilon^{\mu\nu\alpha\beta}
F_{\alpha\beta}$. Indeed, written in this form, one can identically replace all
$\partial_\mu A_\nu$ by $\frac12 F_{\mu\nu}$. Conversely, since the following
identities
\begin{equation}
F^{\mu\nu}F_{\mu\nu} = - \epsilon^{\mu\nu\alpha\beta}
\epsilon_{\rho\sigma\alpha\beta} \partial_\mu A_\nu \partial^\rho A^\sigma \,,
\label{EqLFFEps}
\end{equation}
and
\begin{equation}
\tilde{F}^{\mu\nu} F_{\mu\nu} = 2\epsilon^{\mu\nu\rho\sigma} \partial_\mu A_\nu
\partial_\rho A_\sigma \,,
\label{EqLFtildeFEps}
\end{equation}
hold, any function of $F$ and $\tilde F$ can be expressed as a term such as
discussed above.

This leads to the first Lagrangian compatible with our requirements,
namely the so-called $\mathcal{L}_2$, containing all possible scalars
made by contracting $A_\mu$, $F_{\mu\nu}$ and $\tilde F_{\mu\nu}$. Such
terms can always be expressed \cite{Fleury:2014qfa} as functions of the
scalars $X$, $F^2\equiv F_{\mu\nu} F^{\mu\nu}$, $F\cdot \tilde F
\equiv F_{\mu\nu} \tilde F^{\mu\nu}$ and
\begin{equation}
(A\cdot \tilde F)^2 \equiv A_\alpha \tilde F^{\alpha\sigma} A_\beta \tilde
F^\beta{}_\sigma = A_\alpha A_\beta
\epsilon^{\alpha\sigma\mu\nu} \epsilon^\beta{}_{\sigma\kappa\delta}
\partial_\mu A_\nu \partial^\kappa A^\delta \,,
\end{equation}
again up to integrations by part. The Lagrangian $\mathcal{L}_2$ always
satisfies the conditions discussed in the previous section, and in
particular yields a trivially vanishing Hessian condition
$\mathcal{H}^{0\mu}$: varying $\mathcal{L}_2$ with respect to $\partial
_0 A_0$ [see Eq. \eqref{Hessian}] yields a factor containing
$\epsilon^{00-}$, which vanishes identically. It also gives second order
equation of motion both for $\pi$ and $\bar A_\mu$ as it contains
neither $\partial \partial \pi$ nor $\partial \partial \bar A_\mu$
terms. We should emphasize at this point that $\mathcal{L}_2$ contains
parity conserving as well as parity violating terms; we shall not
consider them any more, but they should be assumed always present in the
forthcoming discussion.

All the terms contained in Eq. \eqref{GenLag} but not of the form discussed
in the previous paragraph read
\begin{equation}
\mathcal{L}^i_N = f^i_N \left( X\right)
\underbrace{\epsilon^{\mu-}\cdots\epsilon^{\nu-} }_N \partial_\mu A_\nu
\cdots \,,
\label{Lgen}
\end{equation}
where at least one field derivative has indices contracted with two
distinct Levi-Civita tensors and the $f^i_N \left( X\right)$ are
arbitrary functions of the gauge vector magnitude $X=A^2$. For $N\leq 2$,
the Hessian condition is automatically satisfied: $\mathcal{H}^{0\mu}$
stems for a variation of the Lagrangian with respect to $\partial_0 A_0$
and $\partial_0 A_\mu$. This demands three equal ``0'' indices
distributed on at most two Levi-Civita tensors, resulting in a vanishing
contribution for symmetry reasons. The other requirements, such
as the order of the equations of motion these terms lead to, are
discussed in length in Secs. \ref{ParCons} and \ref{ParViol}.

For $N>2$, the situation is less clear, as the Hessian does not then
identically vanish. Instead, the condition $\mathcal{H}^{0\mu}=0$ then
implies that the coefficients of all the linearly independent terms
stemming from this condition vanish. The number of such linearly
independent terms increases with the number of field derivatives allowed
for in the Lagrangian, and it is therefore to be expected that, above a
given threshold value $N>N_\mathrm{thr}$, up to unforeseeable fortuitous
cancellations, no new term will be obtainable that could possibly
satisfy the requirements of a safe theory. We conjecture that, as in the
scalar galileon case, $N_\mathrm{thr}=2$; the following sections detail
the reasons hinting to such a conjecture, splitting into parity
conserving ($N$ even) and violating ($N$ odd) contributions. Note that
there exists a general argument, based on the fact that the Lagrangian
contains second-order derivatives of the field, for which the scalar galileon
theory automatically stops at $N=2$ \cite{Deffayet:2009mn}, whereas in
the vector case, no such argument can be found, the Lagrangian
containing only first-order derivatives and there could exist terms
which vanish when $A_\mu\to \partial_\mu \pi$ while still satisfying all
other hypothesis. As a result, the arguments below are different from
those needed to show $N_\mathrm{thr}=2$ in the scalar galileon case.

\section{Parity conserving terms}
\label{ParCons}

Previous works discussed parity conserving actions with $N=2$ including
up to 4 field derivatives $\partial A$, the so-called $\mathcal{L}_n$,
with $n=3,\cdots,5$ \cite{Heisenberg:2014rta}, and $n=6$
\cite{Allys:2015sht}, $n$ counting the number of field derivative plus
two (this convention, bizarre in the vector case, is meaningful in the
original galileon construction). Up to $n=5$, the Lagrangians satisfy
the condition that the scalar part of the vector field corresponds only
to non trivial total derivative interactions, a condition which, once
relaxed, yields the extra $n=6$ term: in the latter situation, one can
always factorize the action by some factors involving the Faraday tensor
and its dual, ensuring it vanishes in the pure scalar sector. All these
terms were shown to be of the form presented in Eq. \eqref{Lgen} above
with $N=2$, thus agreeing with our conjecture. They also comply with all
the necessary requirements we asked for the theory to be physically
meaningful, with second-order equations of motion and only three
propagating degrees of freedom \cite{Allys:2015sht,Jimenez:2016isa}.

In Ref.~\cite{Allys:2015sht}, new terms were also suggested which,
similarly to $\mathcal{L}_6$, were of the form $(\partial A)^p F^q
\tilde F^r$ (with $r$ even to ensure parity conservation), and therefore
vanishing in the pure scalar sector. It was even argued that an infinite
tower of such terms could be generated. A further examination of these
terms however revealed a different, and somehow more satisfactory,
picture: some new terms, by virtue of the Cayley-Hamilton theorem,
vanish identically in 4 dimensions, a conclusion that can also be
reached by rewriting the relevant terms in the form presented in Eq.
\eqref{Lgen}, but with Levi-Civita tensors having more than 4 indices
\cite{Jimenez:2016isa}, explaining why the new terms identically
vanish in 4 dimensions to which the present analysis is restricted: in a
way similar to Lovelock theory for a spin 2 field \cite{Lovelock:1971yv},
one can imagine that
for each number of dimensions, a finite number of new terms can be
generated.

In conclusion of this short section, suffices it to say that extra
parity preserving terms involving more fields and not already present
in $\mathcal{L}_2$ have been actively searched for, and never found;
although this does not prove that such terms cannot be found, this
provides a sufficiently solid basis to assume this statement as a
conjecture, which will only make sense provided a similar conclusion can
be reached for the parity-violating terms to which we now turn.

\section{Parity violating terms}
\label{ParViol}

Parity-violating terms can be written as in Eq. \eqref{Lgen} with an odd
number $N$ of Levi-Civita tensors. For $N=1$, it leads to an action
built from Eq. \eqref{EqLFtildeFEps}, and hence is already included in
$\mathcal{L}_2$ discussed above. One thus expects no terms not included
in $\mathcal{L}_2$ since those terms would contain at least three Levi-Civita symbols. In Ref.~\cite{Allys:2015sht} however, two extra such
terms were found to satisfy all the physically motivated requirements,
obtained through the systematic Hessian method. They read
\begin{equation}
\label{EqL5Eps}
\mathcal{L}_{5}^{\epsilon} = F_{\mu \nu} \tilde{F}^{\mu \nu}
\left(\partial \cdot A \right) - 4  \left( \tilde{F}_{\rho \sigma}  
\partial ^\rho A_\alpha \partial^\alpha A^\sigma \right) \,,
\end{equation}
and
\begin{equation}
\label{EqL6Eps}
\mathcal{L}_{6}^{\epsilon}=  \tilde{F}_{\rho \sigma} F^\rho{}_\beta
F^\sigma{}_\alpha \partial^\alpha A^\beta \,.
\end{equation}
According to our conjecture, they should either vanish or be contained
in the previous terms up to a total derivative. We show below that it is
indeed the case, and for that purpose we first recall an identity derived and first reported, to our knowledge, in 
Ref.~\cite{Fleury:2014qfa}; this completes
the proof that the systematic procedure could not find terms having 
up to 4 field derivatives that are not contained in $\mathcal{L}_2$.

\subsection{A useful identity}

Let $A_{\mu\nu}$ and $B_{\mu\nu}$ be two antisymmetric tensors in a
four-dimensional spacetime with mostly positive signature. One has
\begin{equation}
A^{\mu \alpha} \tilde{B}_{\nu \alpha} + B^{\mu \alpha} \tilde{A}_{\nu
\alpha} = \frac{1}{2} (B^{ \alpha \beta} \tilde{A}_{\alpha \beta}
)\delta^{\mu}_{\nu} \,,
\label{EqFP}
\end{equation}
where $\tilde{X}^{\mu\nu}=
\frac{1}{2}\epsilon^{\mu\nu\alpha\beta}X_{\alpha\beta}$ is the Hodge
dual of $X$ \cite{Fleury:2014qfa}. 

In order to prove this identity, one uses the relation (see, e.g., Ref. \cite{Wald:1984rg})
\begin{equation}
\epsilon^{\alpha_1 \dots \alpha_k \delta_{(k+1)}\dots \delta_n}
\epsilon_{\beta_1 \dots \beta_k \delta_{(k+1)}\dots \delta_n} =
(-1)^s (n-k)! k! \delta^{[\alpha_1\dots}_{\beta_1\dots}\delta^{\alpha_k ]}_{\beta_k} \,,
\label{EqContractionEpsGen}
\end{equation}
where $s$ counts the number of minus signs in the signature of the metric
and $n$  the dimension of spacetime. One gets 
\begin{equation}
\varepsilon^{\alpha_1 \alpha_2 \alpha_3 \delta }\varepsilon_{\beta_1
\beta_2 \beta_3 \delta } = - 3! \delta^{[{\alpha_1 }}_{\beta_1 }
\delta^{\alpha_2}_{\beta_2} \delta^{\alpha_3]}_{\beta_3} \,,
\label{EqContractionEps}
\end{equation}
and
\begin{equation}
\varepsilon^{\alpha_1 \alpha_2 \delta_1 \delta_2 }\varepsilon_{\beta_1
\beta_2 \delta_1 \delta_2 } = - 2! 2! \delta^{[{\alpha_1 }}_{\beta_1 }
\delta^{\alpha_2]}_{\beta_2} \,,
\end{equation}
in the $n=4$-dimensional case, leading to
\begin{equation}
X^{\alpha\beta}=-\frac{1}{2}\epsilon^{\mu\nu\alpha\beta}\tilde{X}_{\mu\nu} \,,
\end{equation}
to express a tensor from its Hodge dual.

Beginning from the left-hand side of the identity we wish to prove, we get
\begin{equation}
A^{\mu\alpha} \tilde{B}_{\nu \alpha} = -\frac14
\varepsilon^{\gamma \epsilon \mu \alpha } 
\varepsilon_{\nu  \sigma \rho \alpha} \tilde{A}_{\gamma \epsilon}
B^{\sigma \rho} \,,
\end{equation}
which, upon using Eq. \eqref{EqContractionEps}, yields 
$A^{\mu \alpha} \tilde{B}_{\nu \alpha} 
= \displaystyle \frac32 \delta^{[{\gamma }}_{\nu }
\delta^{\epsilon}_{\sigma} \delta^{\mu]}_{\rho} 
\tilde{A}_{\gamma \epsilon} B^{\sigma \rho}.$
Expanding and simplifying the relevant terms, one finally
obtains Eq. \eqref{EqFP}, as desired.

As a direct consequence, we can easily deduce the identities
\begin{equation}
F^{\mu \alpha } F_{\nu \alpha}  - \tilde{F}^{\mu \alpha } \tilde{F}_{\nu
\alpha} = \frac{1}{2} \left(F^{ \alpha \beta } {F}_{\alpha \beta}\right)
\delta^{\mu}_{\nu} \,,
\label{EqFPExample1}
\end{equation}
and
\begin{equation}
F^{\mu \alpha } \tilde{F}_{\nu \alpha} = \frac{1}{4} \left(F^{ \alpha \beta }
\tilde{F}_{\alpha \beta}\right) \delta^{\mu}_{\nu} \,,
\label{EqFPExample2}
\end{equation}
which follows from substituting  $A_{\mu\nu}=F_{\mu\nu}$, $B_{\mu\nu}
= \tilde{F}_{\mu\nu}$ and $A_{\mu\nu}=B_{\mu\nu}=F_{\mu\nu}$
respectively in Eq. (\ref{EqFP}).

\subsection{Simplification of $\mathcal{L}_5^{\epsilon}$ and
$\mathcal{L}_6^{\epsilon}$}

We now use the above identities to first expand
$\mathcal{L}_5^{\epsilon}$. One has
\begin{equation}
\mathcal{L}_5^{\epsilon} =  F_{\mu \nu}\tilde{F}^{\mu \nu} \partial
\cdot A - 4 \tilde{F}_{\rho \sigma}  \partial^{\rho} A^{\alpha}
\partial_{\alpha} A^{\sigma}= F_{\mu \nu}\tilde{F}^{\mu \nu} \partial
\cdot A - 4 \tilde{F}_{\rho \sigma}  (F^{\rho \alpha} +\partial^{\alpha}
A^{\rho} )\partial_{\alpha} A^{\sigma} \,,
\end{equation}
whose last term can be transformed into 
\begin{equation}
\tilde{F}_{\rho \sigma}  \left(F^{\rho \alpha}
+\partial^{\alpha} A^{\rho} \right)\partial_{\alpha} A^{\sigma} = 
\frac{1}{4}F\tilde{F} \delta_{\sigma}^{\alpha}  \partial_{\alpha} A^{\sigma} 
+ \tilde{F}_{\rho \sigma}  \partial^{\alpha} A^{\rho} \partial_{\alpha}
A^{\sigma} \,,
\end{equation}
so that, finally, one ends up with 
\begin{equation}
\mathcal{L}_5^{\epsilon} = - 4 \tilde{F}_{\rho \sigma} 
\partial^{\alpha} A^{\rho} \partial_{\alpha} A^{\sigma} = 0 \,,
\end{equation}
being a contraction between a fully symmetric and a fully antisymmetric
tensor. 

As for $\mathcal{L}_6^{\epsilon}$, one has
\begin{equation}
 \mathcal{L}_6^{\epsilon} =  \left(\tilde{F}_{\rho \sigma} F^{\sigma
 \alpha} \right) F^{\rho \beta} \partial_{\alpha} A_{\beta}
 =\left(-\frac{1}{4}F\tilde{F} \delta_{\rho}^{\alpha} \right) F^{\rho
 \beta} \partial_{\alpha} A_{\beta} \,.
\end{equation}
A few straightforward manipulations then yield
\begin{equation}
\mathcal{L}_6^{\epsilon} = -\frac{1}{8} \left(F\tilde{F}\right) F^2 \,,
\label{L6e}
\end{equation} 
showing $\mathcal{L}_6^{\epsilon}$ is not in fact a new term but is already
included in the Lagrangian $\mathcal{L}_2$.

\section{Final model}
\label{PartFinalModel}

The two extra parity-violating terms obtained in
Ref. \cite{Allys:2015sht} have been shown here to be either vanishing or
already included in a previous Lagrangian. As mentioned below
Eq.~\eqref{GenLag}, we also found that the term
\begin{equation}
\mathcal{L}^\mathrm{bis}_4 = g_4(X) A^\mu A_\lambda \tilde F_{\mu\nu}
\partial^\lambda A^\nu = g_4(X) A^\mu A_\lambda \epsilon_{\mu\nu\rho\sigma}
\partial^\rho A^\sigma \partial^\lambda A^\nu ,
\label{FS}
\end{equation}
is compatible with all the conditions we demand and could therefore be
included in the general analysis. The corresponding term in
Eq.~\eqref{GenLag} would be proportional to $\tilde
F_{\mu\nu}S^{\mu\nu}$, with $S^{\mu\nu} = \partial^\mu A^\nu
+\partial^\nu A^\mu$ being the symmetric counterpart of the Faraday tensor which
clearly vanishes identically. The line of reasoning leading to
a small number of possible terms of the form \eqref{GenLag} should apply
to higher order terms of the kind \eqref{FS}; we did not find any such
terms.

According to all the above discussions, it seems safe to conjecture that the final
complete action for a Proca vector field involving only first-order
derivatives in 4 dimensions is that given by Eq.~(12) of
Ref. \cite{Jimenez:2016isa}, together with Eq.~\eqref{FS}.
The complete formulation of the
parity-conserving terms were also derived and written in a simple form
in Refs. \cite{Heisenberg:2014rta,Allys:2015sht,Jimenez:2016isa}. We merely
repeat the full action below:
\begin{equation}
\mathcal{S} = \int\dd^4 x\sqrt{-g} \left( -\frac{1}{4} F_{\mu\nu}
F^{\mu\nu} + \sum_{i=2}^6 \mathcal{L}_i + \mathcal{L}^\mathrm{bis}_4  \right),
\label{TotalLag}
\end{equation}
with 
\begin{equation}
\begin{split}
\mathcal{L}_2 &= f_2\left( A_\mu, F_{\mu\nu}, \tilde{F}_{\mu\nu} \right) = f_2\left[
X,F^2,F\cdot \tilde F,\left( A \cdot \tilde F\right)^2\right] \,, \\
\mathcal{L}_3 &= f_3\left( X\right) \partial \cdot A = \frac12 f_3\left( X\right) S_{\mu}{}^{\mu} \,, \\
\mathcal{L}_4 &= f_4\left( X\right) \left[ (\partial \cdot A)^2-\partial_\rho
A_\sigma \partial^\sigma A^\rho \right] = \frac14 f_4 \left( X\right) \left\{ \left[ (S_{\mu}{}^{\mu})^2-S_{\rho}{}^{\sigma} S_{\sigma}{}^{\rho} \right] + F_{\mu\nu}F^{\mu\nu}\right\} \,, \\
\mathcal{L}_5 &= f_5\left( X\right)\left[ (\partial \cdot A)^3-3 (\partial \cdot A)
\partial_\rho A_\sigma \partial^\sigma A^\rho + 2 \partial_\rho A^\sigma
\partial_\gamma A^\rho \partial_\sigma A^\gamma\right] + g_5\left( X\right)
\tilde{F}^{\alpha\mu} \tilde{F}^\beta{}_\mu \partial_\alpha A_\beta \\
& = \frac18 f_5\left( X\right)\left[ (S_{\mu}{}^{\mu})^3-3 (S_{\mu}{}^{\mu}) 
S_{\rho}{}^{\sigma}S_{\sigma}{}^{\rho}  + 2 S_{\rho}{}^{\sigma}S_{\sigma}{}^{\gamma}
S_{\gamma}{}^{\rho}  \right] + \frac14 \left[ 2 g_5\left( X\right) - 3 f_5\left( X\right)\right]
\tilde{F}^{\alpha\mu} \tilde{F}^\beta{}_\mu S_{\alpha \beta } \,, \\
\mathcal{L}_6 &= g_6 \left( X\right) \tilde{F}^{\alpha\beta} \tilde{F}^{\mu\nu} \partial_\alpha 
A_\mu \partial_\beta A_\nu =\frac14 g_6 \left( X\right) \tilde{F}^{\alpha\beta} \tilde{F}^{\mu\nu}
\left( S_{\alpha\mu} S_{\beta\nu} + F_{\alpha\mu} F_{\beta\nu}\right).
\end{split}
\label{Lterms}
\end{equation}
In Eq.~\eqref{Lterms}, $f_2$ is an arbitrary function of all possible scalars made out of
$A_\mu$, $F_{\mu\nu}$ and $\tilde F_{\mu\nu}$, containing both parity
violating and preserving terms, while $f_3$, $f_4$,  $f_5$, $g_5$ and
$g_6$ are arbitrary functions of $X$ only. Note that this dependence
is compatible with our basis choice in Eq.~\eqref{GenLag}, so that any other choice,
for instance $g_k (X, F^2)$, would spoil the Hessian condition. 
We assume also that the standard kinetic term, $-\frac14 F_{\mu\nu} F^{\mu\nu}$, 
does not appear
in $f_2$, in order that the normalization of the vector field follows
that of standard electromagnetism and thus we have pushed it out in
Eq.~\eqref{TotalLag}. The Lagrangians of Eq. \eqref{Lterms} are expressed in
terms of either the ordinary derivatives $\partial_{\mu} A_{\nu}$, or in
terms of its symmetric $S_{\mu\nu}$ and antisymmetric $F_{\mu\nu}$
parts\footnote{The relation between our formulations and those in terms
of the Levi-Civita tensors is given in Ref.~\cite{Jimenez:2016isa}.}: the
second formulation, obtained by setting $\partial_{\mu} A_{\nu}=\frac12
\left( S_{\mu\nu} + F_{\mu\nu}\right)$ and making use, in the case of
$\mathcal{L}_5 $, of Eq.~\eqref{EqFPExample1}, induces extra terms in
$\mathcal{L}_4$ and $\mathcal{L}_6$ which can be absorbed in the
parity-preserving part of $\mathcal{L}_2$, being functions of $A_\mu$
and $F_{\mu\nu}$.

The presence of the new term $\mathcal{L}^\mathrm{bis}_4 $ is not as
surprising as it would appear at first sight when one considers the
generic structure of the terms contained in Eq.~\eqref{Lterms}. For the
dynamics of the Lagrangians to be non trivial, up to terms already
contained in $\mathcal{L}_2$, the functions $f_3$, $f_4$, $f_5$ and
$g_6$ must contain at least one factor of $X=g_{\mu\nu}A^\mu A^\nu$ (see
also Ref.~\cite{Jimenez:2016isa}). Assuming $2g_5 - 3 f_5$ to also
contain such a factor (generic situation, no fine-tuning of the
arbitrary functions), we conclude that each term can be written in the
form\footnote{Sometimes also up to terms included in $\mathcal{L}_2$.}
\begin{equation}
\mathcal{L}_i = A^2 h(X) \langle \mathcal{O}_i\rangle = \tilde h(X) \langle A
\tilde{\mathcal{O}}_i A \rangle + \partial_\mu J_i^\mu,
\label{Li}
\end{equation}
$A^2 h$ standing for the relevant $f$ or $g$ function (this
transformation is indeed not possible for $\mathcal{L}^\mathrm{bis}_4$). 
In Eq.~\eqref{Li},
the brackets indicate a trace over spacetime indices,
$\mathcal{O}_i$ and $\tilde{\mathcal{O}}_i$ are operators constructed
from $\tilde F$'s (possibly none) and at least one $S$, and $J_i^\mu$ is the relevant
current to make the identity true\footnote{See
Ref.~\cite{Deffayet:2011gz} for an extensive discussion of these
equivalent formulations in the scalar galileon case.}. So, the terms
vanishing in the purely scalar case, i.e. those for which
$\mathcal{O}_i$ contains at least one factor of $\tilde F$, take the
form $\langle AS\tilde FA\rangle$, $\langle AS\tilde F\tilde F A\rangle$
and $\langle AS\tilde FS\tilde F A\rangle$. The first such term, which
is nothing but our $\mathcal{L}^\mathrm{bis}_4 $, is then seen to appear
in a totally natural way.

Our final action is, up to the new term $\mathcal{L}^\mathrm{bis}_4$,
exactly the same as that of Ref.~\cite{Jimenez:2016isa}. There is
however a subtle difference in the fact that all possible
parity-violating terms are also written, being included in $f_2$ and
$\mathcal{L}^\mathrm{bis}_4$. Note that the curved space-time
generalization of this action is also given in
Ref.~\cite{Jimenez:2016isa}, the covariantization of
$\mathcal{L}^\mathrm{bis}_4$ being obtained by a trivial replacement
$\partial \to \nabla$.

A legitimate question to ask is whether Eq. \eqref{TotalLag} is indeed
the most general theory that can be written involving a vector field
with three propagating degrees of freedom and second-order equations of
motion. This has already been conjectured in
Refs.~\cite{Heisenberg:2014rta,Jimenez:2016isa}. Now, the discussion and
calculations of the present article correct the conjecture made in Ref.
\cite{Allys:2015sht} about an infinite tower of terms, and also suggests
a finite number of terms, even in the parity violating sector. So, there
is finally a complete agreement on this point.

An additional indication of the correctness of this conjecture is that
the systematic investigation procedure of Ref. \cite{Allys:2015sht}
completed by the calculation of Ref. \cite{Jimenez:2016isa} for the
parity-conserving sector, and by the present paper in the
parity-violating sector, did not find any term other than those shown
above up to the orders of $\mathcal{L}_6$ (parity violating) and
$\mathcal{L}_7$ (parity conserving). However, if there were an infinite
tower of possible Lagrangians, one would expect such Lagrangians to
appear in our systematic procedure, which is not the case. Note
especially that these works show that the parity violating sector
contains no other terms than $\mathcal{L}^\mathrm{bis}_4$ and those
contained in $f_2$, which is a very strong constraint, and greatly
strengthens the conjecture we have made.

Finally, this work heavily relies on the postulate that spacetime is 4
dimensional. Relaxing this assumption permits to include the extra terms
proposed in Ref. \cite{Allys:2015sht} which, as shown in Ref.
\cite{Jimenez:2016isa}, can be expressed with higher dimensional
Levi-Civita tensors. For a given spacetime dimensionality, one thus
expects, just like in the Lovelock case for a spin 2 field \cite{Lovelock:1971yv}, a finite
number of new terms to appear: in practice, in $D$ dimensions, one
expects terms containing up to $D$ first order derivatives of the 
vector field.

\section*{Acknowledgments}

We acknowledge P. Fleury and C. Pitrou for pointing out the importance
of the identity in Eq. (\ref{EqFP}). We also  wish to thank C.~Deffayet,
G.~Esposito-Farese, L.~Heisenberg and J.~Beltr\'an Jim\'enez for useful
clarifications and enlightening discussions. This work was supported by
COLCIENCIAS grant numbers 110656933958 RC 0384-2013 and 123365843539 RC
FP44842-081-2014. P.P. would like to thank the Labex Institut Lagrange
de Paris (reference ANR-10-LABX-63) part of the Idex SUPER, within which
this work has been partly done.

\bibliographystyle{unsrt} 
\bibliography{bibli} 

\end{document}